\documentclass{PoS}

\title{Status of the Davies-Cotton and Schwarzschild-Couder Medium-Sized Telescopes for the Cherenkov Telescope Array}

\ShortTitle{Medium-Sized Telescopes for the Cherenkov Telescope Array}

\author{\speaker{J-F. Glicenstein} for the CTA MST and CTA SCT Projects\footnote{for consortium list see PoS(ICRC2019)1177}\\
 IRFU, CEA-Paris Saclay, Universit\'e de Paris-Saclay\\
 F-91191 Gif-sur-Yvette, France 	\\
        E-mail: \email{glicens@cea.fr}
}

\abstract{The Cherenkov Telescope Array is an observatory dedicated to very high energy gamma rays with unprecedented sensitivity between 20 GeV and 300 TeV to be installed on two sites: Canary Island La Palma and Paranal Chile. Three telescope sizes will be used to cover the entire energy range. Two different telescope designs have been proposed to cover the intermediate energy range from 150 GeV to 5 TeV. One of the proposals is based on the traditional single mirror Davies-Cotton design, with two different camera concepts with different detection and processing schemes are under test. Another innovative design based on a dual-mirror Schwarzschild-Couder optics has also been developed and is under test in Arizona, USA. In this talk, the different concepts and the status of the prototypes will be presented.}

\FullConference{36th International Cosmic Ray Conference -ICRC2019-\\
		July 24th - August 1st, 2019\\
		Madison, WI, U.S.A.}

\begin{document}

\section{Introduction}
The Cherenkov Telescope Array is dedicated to very high energy gamma rays in the range 20 GeV to 300 TeV. Several widely different telescope designs have been proposed to cover the
central range in energies, between $\sim 150$ GeV and $\sim 5$ TeV, the so-called Medium-Sized Telescope's range. A first solution, which uses a modified single mirror Davies-Cotton (DC) layout 
inspired by the present generation of Imaging Atmospheric  Cherenkov Telescope (IACT) structures, is described  in Section \ref{sec:mst-str}.  The second telescope design is a dual-mirror 
Schwarschild-Couder (SC) structure, presented in Section \ref{sec:sct-str}. 
 Two different cameras concepts have been designed to equip the focal planes
of DC telescopes: FlashCam (Section \ref{sec:mst-fls}) and NectarCAM (Section \ref{sec:mst-nec}). The SC telescope is equipped with a Silicon Photomultiplier (SiPM) based camera (Section \ref{sec:sct-chech}) placed between the primary and secondary mirrors. 

Medium-Sized telescope structures must verify certain requirements for acceptance by the CTA observatory. In particular
(a) they must be able to retrieve the coordinates of any pointed direction with a precision of 20 arcseconds (7 arcseconds in certain directions),
(b) the field of view of the camera should be $> 7^{\mathrm{o}},$
(c) the angular pixel pitch is less than $0.18 ^{\mathrm{o}}$ to match the optical point spread function (PSF) of the telescope and to fulfill physics requirements on the angular resolution for gamma rays.
The challenge to solve is thus to have  large fields of view with relatively small PSF and a pointing precision better by a factor of 2 than the present generation of Cherenkov telescopes.

\section{Telescope structures}
\subsection{DC structure design and prototyping}\label{sec:mst-str}
\begin{figure}[h]
\centering
\includegraphics[height=4.5cm]{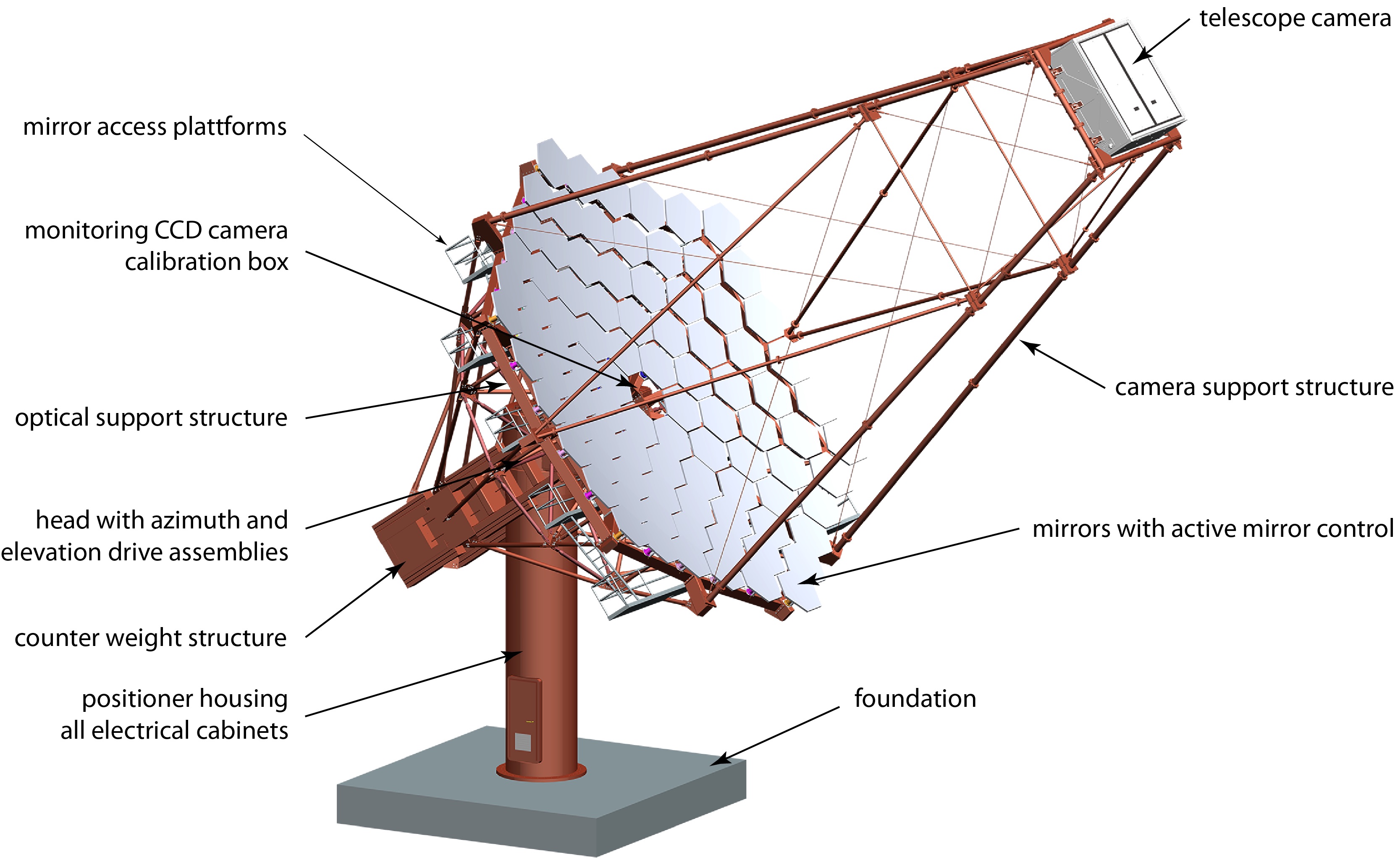}\includegraphics[height=4.5cm]{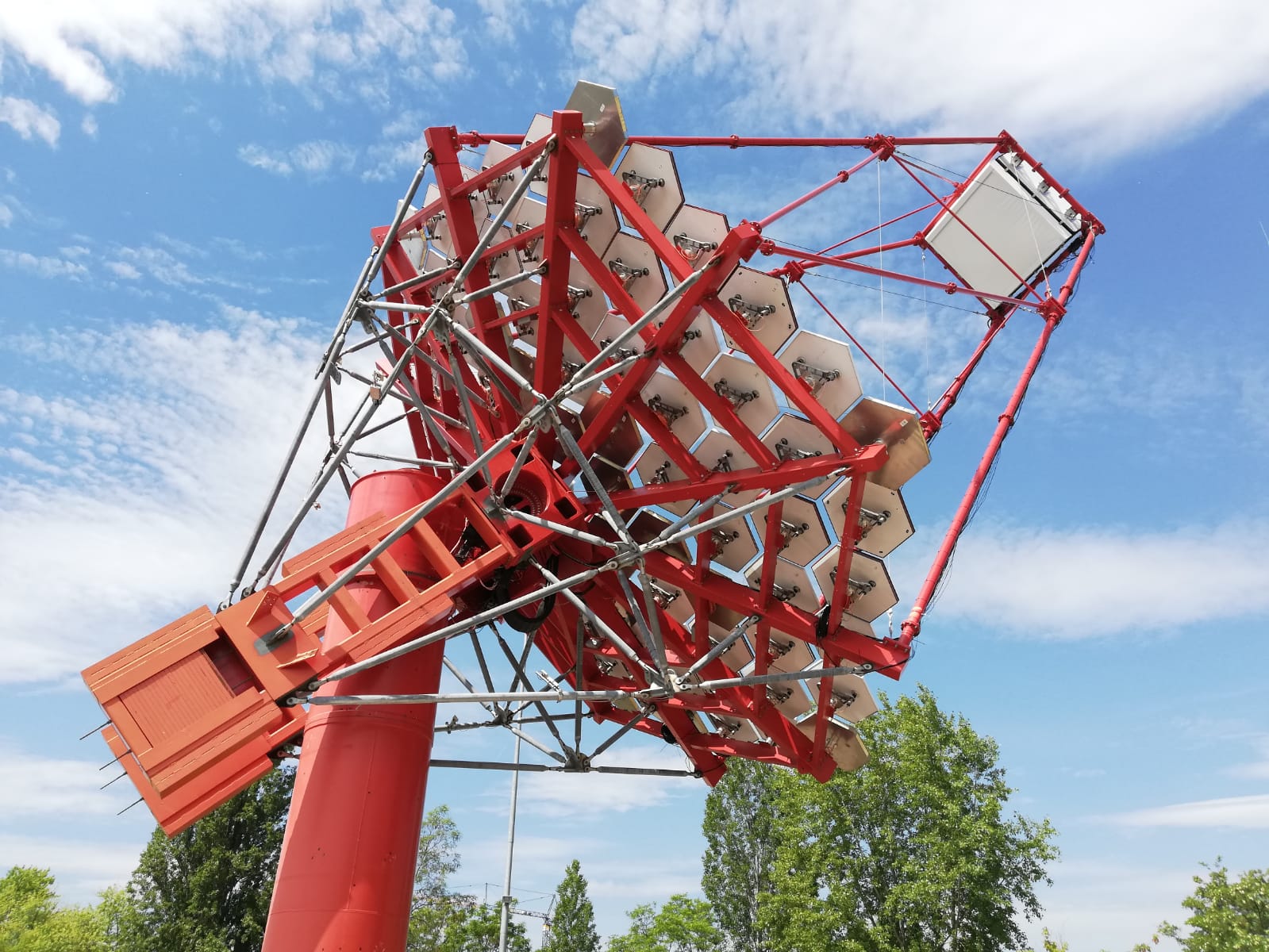}
\caption{Left: Schematics of MST-structure. Right: NectarCAM prototype mounted on the Adlershof structure prototype.}
\label{fig:mst}
\end{figure}

The DC structure layout \cite{bib:Davies}, which has been adopted by the majority of present day Cherenkov telescopes is illustrated in the left panel of Figure \ref{fig:mst}. 
The reflector is a telesselated dish consisting of 86 hexagonally-shaped mirrors
with a 1.2 m flat-to-flat side length. The focal length of the telescope is 16.0 m. Each individual mirror has a nominal radius of curvature of 32.14 m, approximately twice the focal length, 
and they are aligned so that rays parallel to the optical axis converge at the focal point.  
The spherical reflector does not result in synchronous image at the camera plane.
Parabolic mounts do have a
synchronous image, but also larger abberations. 
The dish shape is slightly
distorted to improve the isochronicity of the reflector while keeping the abberations of the 
PSF at a reasonnable level. 

The individual mirrors of the reflector have very stringent requirements on reflectivity (over 85 \% over the whole optical passband)
and their durability.  Several mirror designs, all based on cold slumping technology, have been proposed. 
The mirror spherical shape is produced by bending of a thin glass sheet on a precisely formed
mould. 
Reinforcement of the mirror is made using an aluminium honeycomb sandwich panel glued to the glass
and adding a second piece of glass on the other side of the honeycomb. 
The radius of curvature obtained by the cold slumping technology are within a few centimeters from their nominal value. The spot size at the nominal value is required 
to be less than 12 mm to avoid degrading the telescope PSF.
The telescope structure is not 
protected by a dome as it is usually the case for optical telescopes. The mirror facets have to survive  
in a harsh environment with rain or sand abrasion. To avoid rapid degradation, 
the facets are coated with a protective layer made with a multilayer of SiO2 and HfO2 or ZrO2. 
The facets are aligned 
using actuators during the initial telescope assembly.  The optical support structure is stiff enough so that mirrors
do not need an active alignment during observations.

The other major elements of the structure are the positioner and the camera support structure. 
The positioner of the telescope is designed as a cylindrical tower with
2m diameter and 9m height. The tower is separated into three floors and hosts electrical cabinets, the azimuth
drive and the lubrication assembly. The elevation drive is mounted on the two sides of the head, outside of the
positioner. 
The tubes of the camera support structure are designed to be stiff enough to reduce the flexion of the telescope while  minimizing the shadowing of the mirrors. 
The effective mirror area taking shadowing into account is $\sim 90\ \mathrm{m}^2.$
Since the telescope structure is bent by gravity, the true pointing position is obtained by
comparing the measured optical axis direction for a given (Alt, Az) input to the positions of stars.
The MST pointing calibration
uses a single, wide FoV CCD camera (SingleCCD) installed in the center of the dish, aligned to the optical
axis of the telescope and facing the Cherenkov camera. The FoV of this camera is chosen sufficiently large to
observe both the star field (to determine the direction of the optical axis of the SingleCCD) and the pointing
LEDs mounted on the Cherenkov camera body (to determine its orientation w.r.t. the optical axis).

The design of the MST structure has been studied with successive prototypes. The latest, built in Berlin-Adlershof, is 
 shown in the right panel of Figure \ref{fig:mst}. The prototypes were used to gain experience with the design, leading to
 improvements of the dish and camera support structure.
They have also been used to develop and debug
the telescope and array control software.

A further use of the Adlershof prototype structure is the study of long term behavior of mirrors exposed to daily temperature cycling 
and harsh environmental conditions (wind, rain). Automatic mirror alignment routines  using either an artificial light source 
located on the top of a building at Humboldt University or stars have been developped.  The operation of the 
SingleCCD camera allowed the implementation of a bending model to compensate for telescope flexions.   

The two camera designs which are proposed to equip the focal plane of DC MSTs have different internal architectures, digitization and trigger concepts. However,
both have in common a photon detection based on high quantum efficiency photomultipliers (PMTs), with an associated embarked readout and trigger electronics. 
Both cameras have a field of view of $> 7.5^{\mathrm{o}}$ and roughly 1800 pixels arranged in 
a hexagonal pattern.  The interface to the camera support structure and the power, liquid cooling, dry air, optical fibers for control and data transfer are identical
for both camera types.  The interfaces of cameras to the telescope structure were tested during the FlashCam and NectarCAM integration-test campaign at Adlershof in 2017 and 2019.

\subsection{SC structure design and prototyping}\label{sec:sct-str}
SC telescope (SCT) structures \cite{bib:Couder}  employ a novel aplanatic optical system composed of two
aspheric mirrors
\cite{bib:Vassiliev}.  
The resulting dual-mirror concept has
several potential advantages compared to DC structures:
(a) it corrects for spherical and comatic abberations over a wide field of view, leading to an improved PSF,
(b) the overall camera size is smaller, making possible the use of solid state devices such as  Silicon photomultiplers (SiPM) as active elements,
(c) the smaller-sized camera is expected to be cheaper than the camera for DC structures.

\begin{figure}[h]
\centering
\includegraphics[height=4.5cm]{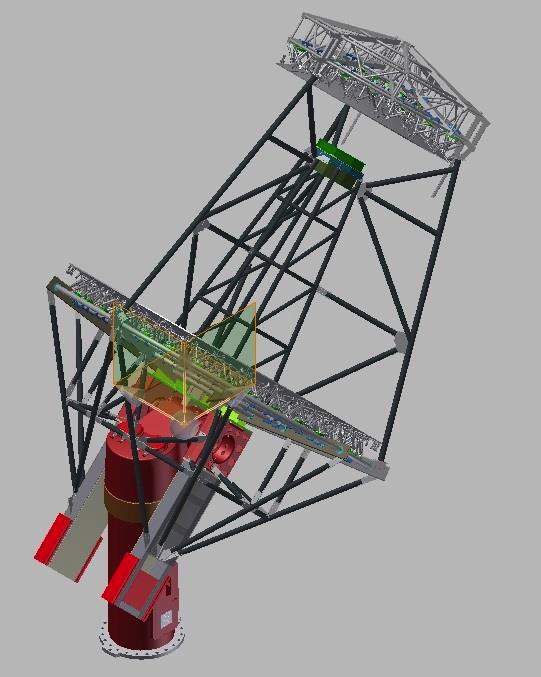}\includegraphics[height=4.5cm]{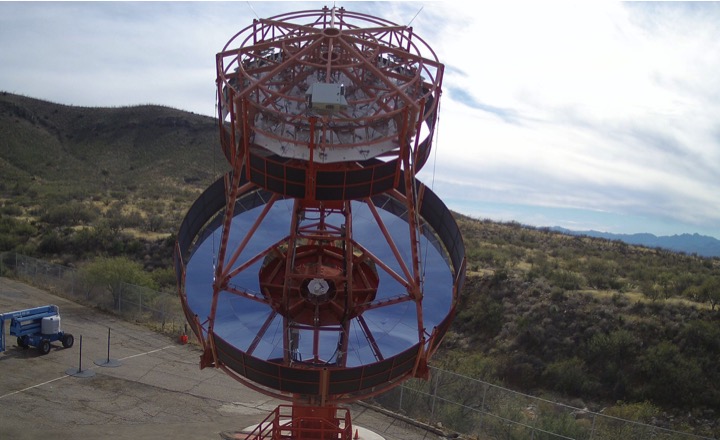}
\caption{Left: Simplified SCT layout, right: SCT prototype installed at the Fred Lawrence Whipple Observatory, showing the pimary mirror}
\label{fig:psct1}
\end{figure}

The SCT structure, shown on the left panel of Figure \ref{fig:psct1}, has a primary mirror of 9.7 m aperture, a secondary mirror of 5.4 m aperture, and a focal
length of 5.6 m ($f/D = 0.58$). The telescope's steel optical support structure (OSS)  supports its camera, mirrors
and auxiliary systems, and is mounted to a main plate, along with a counterweight structure, onto the elevation
axis of a positioner composed of a head and a tower. The SCT's positioner is very similar to that constructed
for the prototype Medium-Sized telescope with DC design (Section \ref{sec:mst-str}), with differences such as a reduced
height of the tower.  
The SCT mirror surface  consists in 72 mirror panels of 4 different shapes. The primary
mirror surface contains an inner ring of 16 panels, and an outer ring of 32 panels. The secondary mirror surface
consists of an inner ring of 8 panels, and an outer ring of 24 panels. The total mirror area is $\sim 50 \mathrm{m}^2,$
with a shadowing of $<12$\%. 
The mirrors are aligned by 444 actuators.  Mirror alignment is one of the major challenges of the SCT. In 
order to have a PSF of the order of the expected size of the camera pixel, the mirror alignment must be performed at the sub-mm level.  

The proof-of-concept,
medium-sized Schwarzchild-Couder telescop (pSCT) is shown on the right panel of
Figure \ref{fig:psct1}. 
The pSCT's positioner was successfully installed in February 2016, and
the successful assembly and erection of the major components of the OSS was completed by August 2016.
The pSCT has a camera access tower which was installed in April 2019. 
The metrology of the primary and
secondary dishes was measured after their installation on the telescope structure and is within specifications.
The alignment of the optical system is undergoing. The pSCT was inaugurated in January 2019 and had its first light
early 2019 (Right panel of Fig. \ref{fig:shower}).

\section{Cameras}
The cameras for the Medium-Sized telescopes of CTA have many stringent requirements including:  
(a) the event timing for large signals should have a precision better than 2 ns,
(b) the full waveform of triggered events should be recorded during 60 ns,
(c) the dynamic range of each pixel is 0 to 2000 photoelectrons (p.e),
(d) the instrument deadtime must be $<5\%,$
(e) the gain calibration should have a precision of less than $8\%.$

\subsection{Camera concept for the DC structure: FlashCam}\label{sec:mst-fls}

The FlashCam camera \cite{bib:Werner} is based on a fully digital trigger and read-out system. The front part is composed 
of a photon detection plane (PDP) comprising 147 modules with 12 PMT each. Each PDP module provides the high voltage for photomultipliers, a pre-amplifier and   an interface for
slow control, monitoring, and safety functions. 
\begin{figure}[h]
\centering
\includegraphics[height=4.5cm]{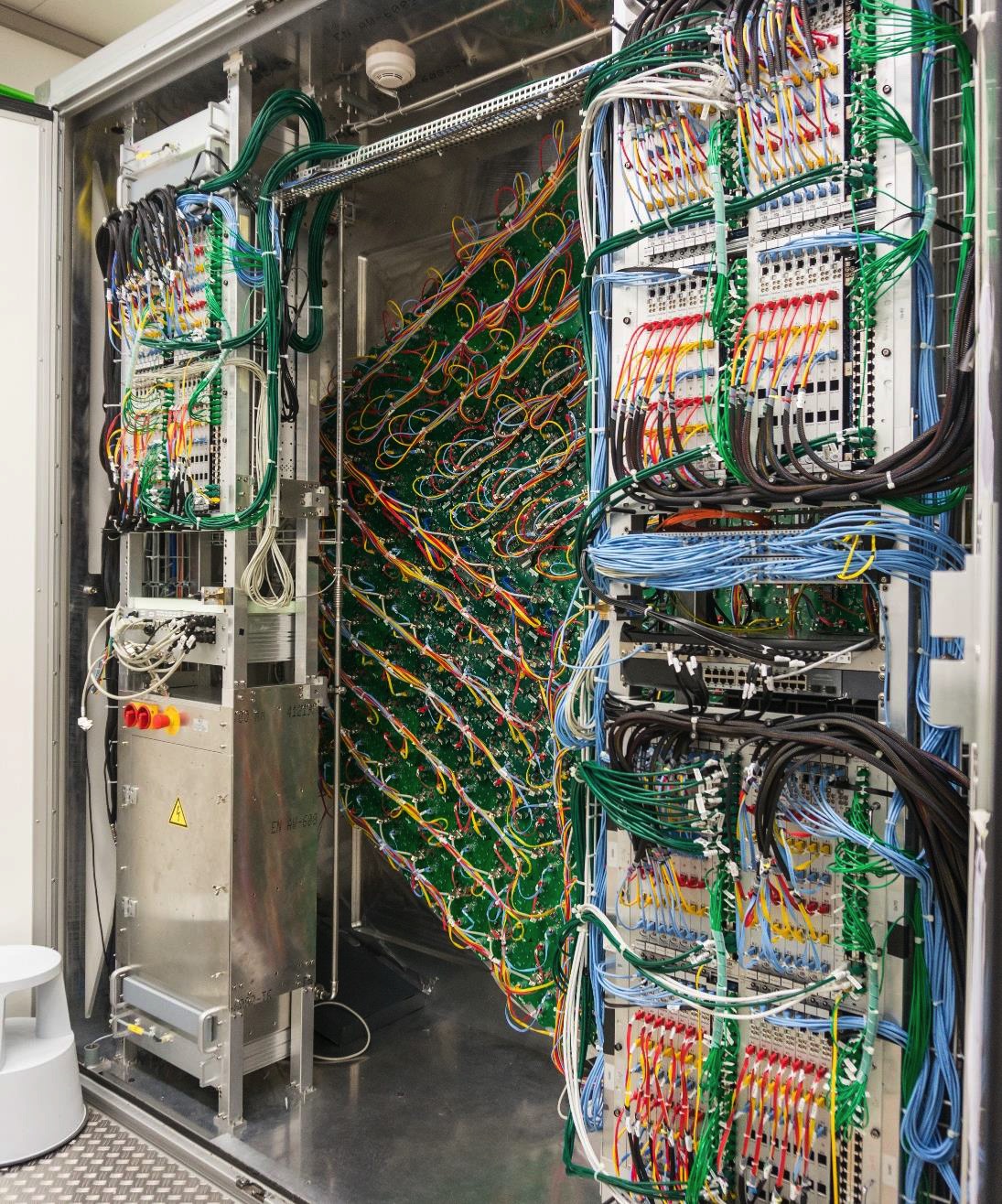}\includegraphics[height=4.5cm]{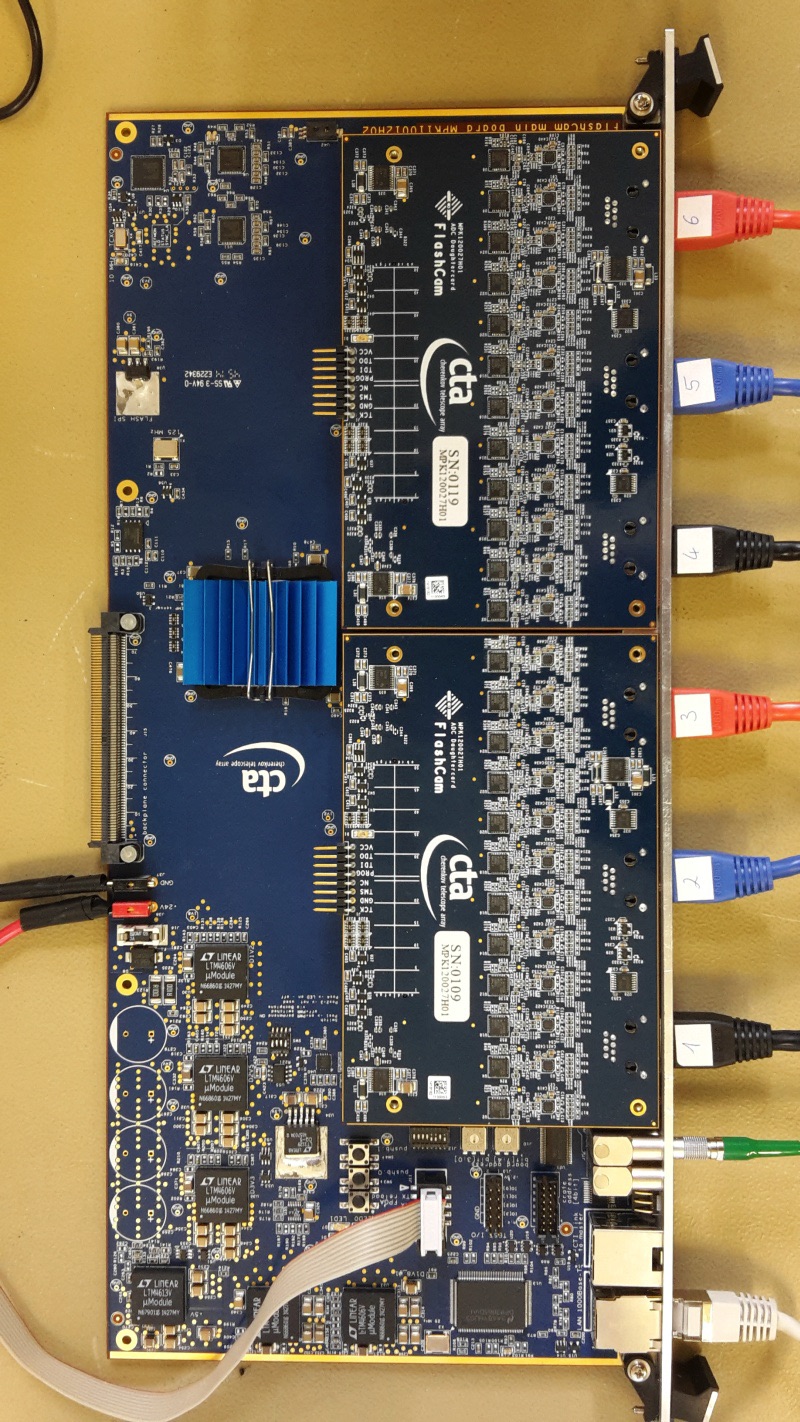}
\caption{Left: Back of FlashCam camera prototype. Right: FlashCam FADC board.}
\label{fig:fcam}
\end{figure} 
A non-linear amplification
scheme is used, in which signal amplitudes larger than about 250 p.e. 
 saturate with the integral growing logarithmically
with input charge. This serves to extend the dynamic range up to beyond 3000 p.e.
while retaining linearity and sub-p.e. resolution for small signals. FlashCam uses only one gain channel, in contrast to NectarCAM,
reducing the required bandwidth of the subsequent data transfer.

The readout electronics boards (serving 24 channels each), as well as
trigger and master distribution boards are organized in crates and racks at the back of the camera, shown in the left panel of Fig.  \ref{fig:fcam}.
The data signal is digitized into 12-bit FADCs (Right panel of Fig. \ref{fig:fcam}), using commercial low-power 250MHz pipeline-ADCs. The
trigger evaluation is performed directly on the data. The trigger creation as well as further pre-processing of the digitized signals is performed in the front-end electronics
by commercial low-cost FPGA.  
At trigger time, waveforms of time slices up to 15.6 $\mu$s long are sent to a camera server using four 10-Gbit Ethernet fibers.  Two additional fibers are used for slow control
monitoring and clock distribution. 

The racks can be accessed for installation and maintenance from behind the camera after opening the rear doors. Photon
detector plane modules are installed from the inside of the camera,
without the need to remove the optical front system. 

\begin{figure}
\centering
\includegraphics[height=4cm]{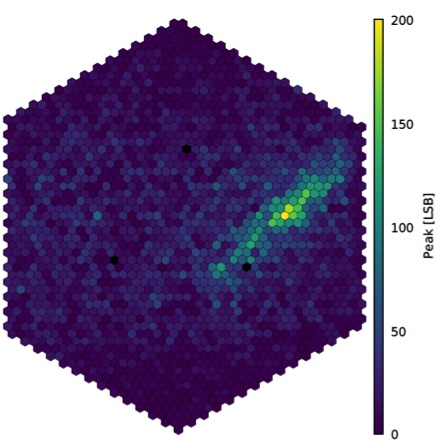}\includegraphics[height=4cm]{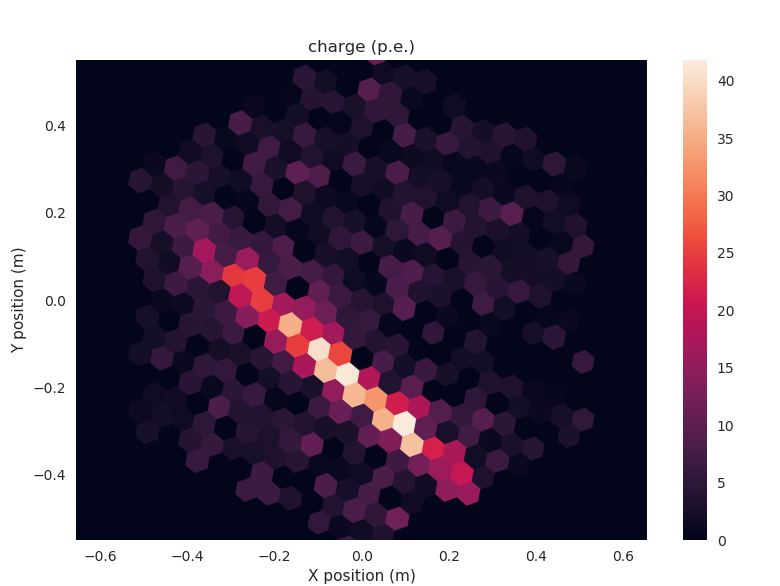}\includegraphics[height=4cm]{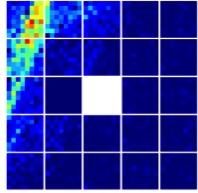}
\caption{Left: Shower event obtained during the FlashCam campaign  at Adlershof in 2017. Center: Shower event obtained during the NectarCAM campaign at Adlershof in 2019.
Right: Shower event obtained with the SCT prototype early 2019.}
\label{fig:shower}
\end{figure}

A fully equipped, full-size prototype camera has been
used for the verification of performance parameters such as time and charge resolution. The event capture and transfer 
has been shown to be deadtime free up to 35 kHz. Finally, a fully equipped FlashCam prototype had its first light during a campaign at Adlershof in 2017 (left part
of Fig. \ref{fig:shower}).

\subsection{Camera concept for the DC structure: NectarCAM}\label{sec:mst-nec}
\begin{figure}[h]
\centering
\includegraphics[height=4.5cm]{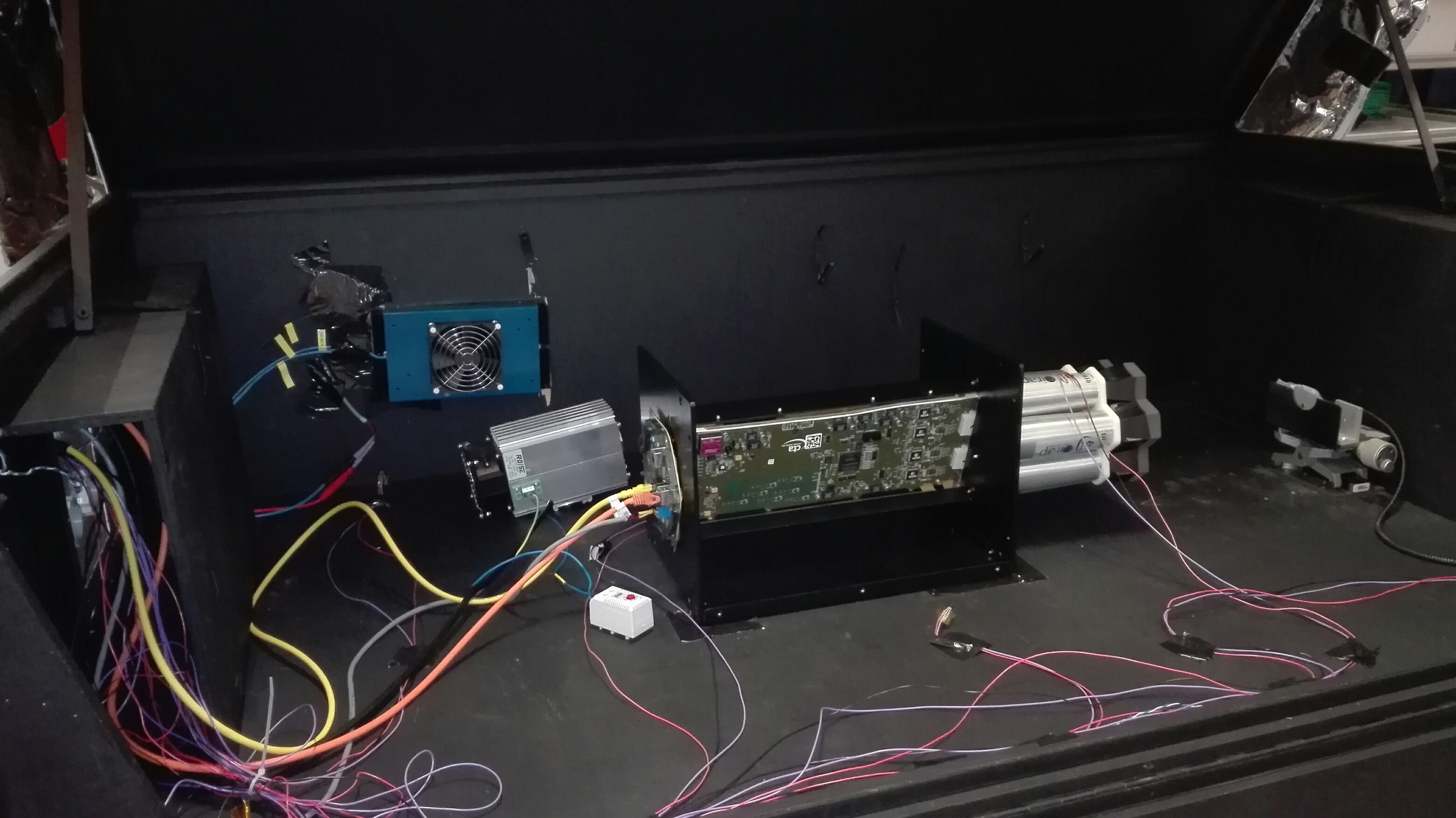}\includegraphics[height=4.5cm]{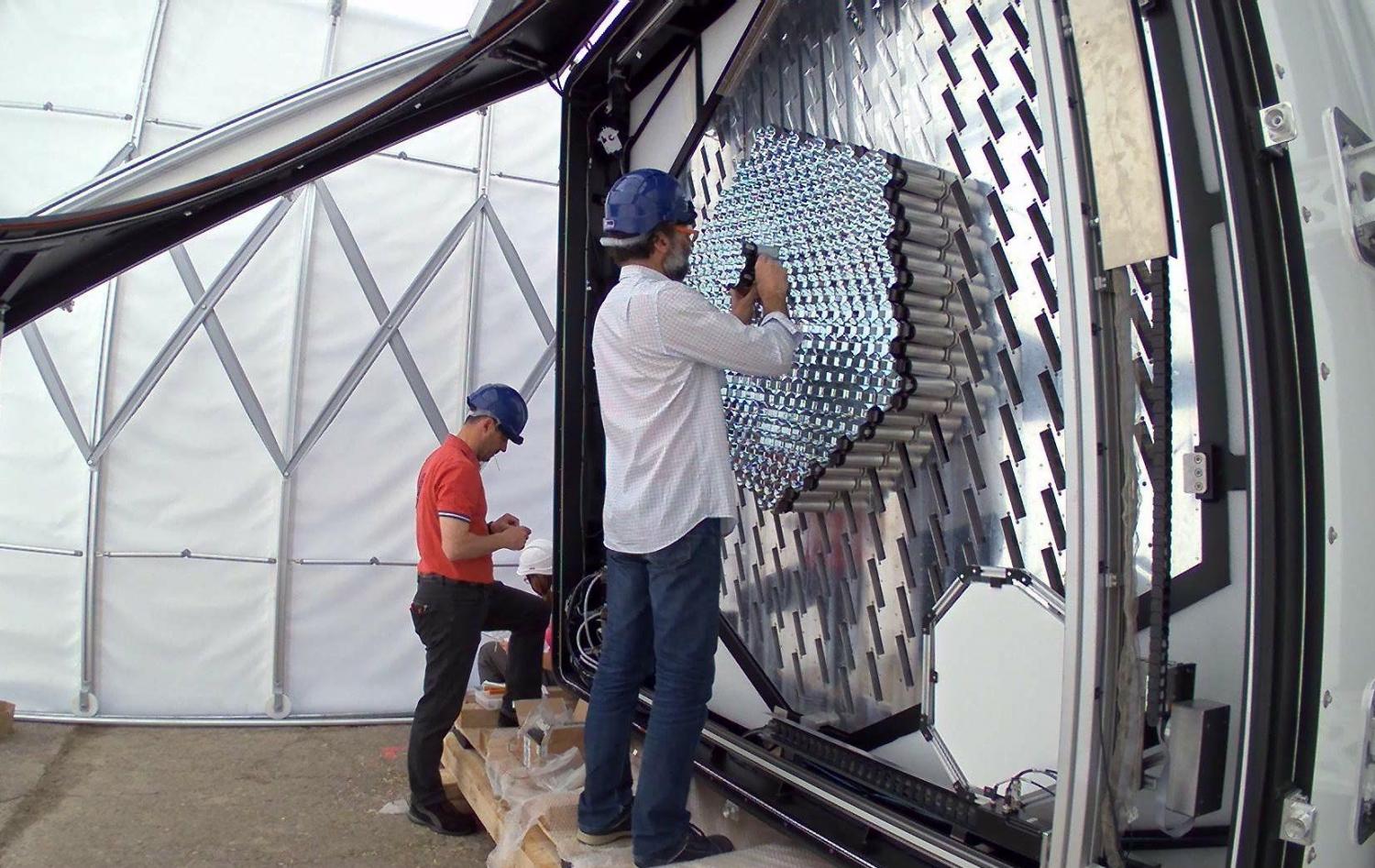}
\caption{Left: Nectar module test bench. Right: NectarCAM prototype being installed at in a tunnel during Adlershof tests.}
\label{fig:nect1}
\end{figure} 
NectarCAM detects photon signals with focal plane modules (FPM). The FPM are composed of 7 PMTs with an associated high voltage and a first level of preamplification board and a control
board with allows to set high voltages and monitor anode currents.  The output of FPM is sent to a front-end board where the signal is amplified a second time and divided into a high gain, 
a low gain and a trigger channel. The two gains are needed to obtain the full 0.5-2000 p.e. dynamic range. Data from high and low gain channels are sampled continuously  at 
an adjustable rate (typically 1 GHz) in Nectar chips. The analogue bandwidth of the data channel 
is typically 300 MHz. Nectar chips integrate both  a $1 \mu$s wide switched capacitor array (SCA) and a 12-bit (effective 11.2-bit) ADC in a single chip. 
The trigger logic is perfomed by sending discriminated signals into a backplane 
and looking for a cluster in 7 neighboring pixels. The trigger are time-stamped with a custom board built on the top of a White Rabbit client.
NectarCAM is modular and contains 265 modules with a FPM, the associated front-end board and backplane. The total field of view is $8^{\mathrm{o}}.$
The dead-time of NectarCAM is dominated by the readout of Nectar chips.  A low dead-time of <5 \% at 7 kHz event rate for a
full camera has been achieved.
For each event, time slices of up to 60 ns are sent to a camera server which performs the event building, online calibration and is located outside the camera. The data from individual modules are first sent 
to a stack of commercial switches located in a crate in the rear of the camera, then to the camera server
by 4\ 10-Gbit fibers.  
 The full data acquisition chain has been tested successfully
at rates up to 15 kHz. 
The PMTs and associated readout modules are installed (and exchanged in case of maintenance
repair) from the front side of the camera.  The total heat dissipation
in the camera is expected to be $\sim 7$ kW and is removed by 2 loops of forced air circulation connected 
to an external water cooling pipe through an heat exchanger.

Components, including full modules (left part of Fig. \ref{fig:nect1}) have been extensively tested and verifed for their performances. 
A camera including the full size camera housing, module holder, entrance window, cooling system, but partially equipped by 
only one quarter of the modules has been assembled early 2019 (right part of Fig. \ref{fig:nect1}) . Its performances have been extensively tested inside a dark room at the    
 NectarCAM camera assembly and integration site. More details on the NectarCAM camera test results are given in \cite{bib:Tavernier}. 
 The mini-camera fulfills all CTA scientific and technical requirements. The partially equipped camera has been mounted on the MST structure prototype and had its first light 
 in May 2019 (central panel of Fig. \ref{fig:shower}).

\subsection{Camera concept for the SC structure}\label{sec:sct-chech}
The camera of the SCT is placed at the focus of the secondary mirror, between the primary and the secondary
mirrors. Linear stages mounted to the frame of the camera allow positioning the camera with a precision of 200 $\mu$m
along all three axes, which is a necessary requirement to achieve the design goal for the optical point-spread
function. The photon sensitive area (focal plane) of the camera has a diameter of  0.8m and covers a field
of view of $8^{o}.$

\begin{figure}[h]
\centering
\includegraphics[height=4.5cm]{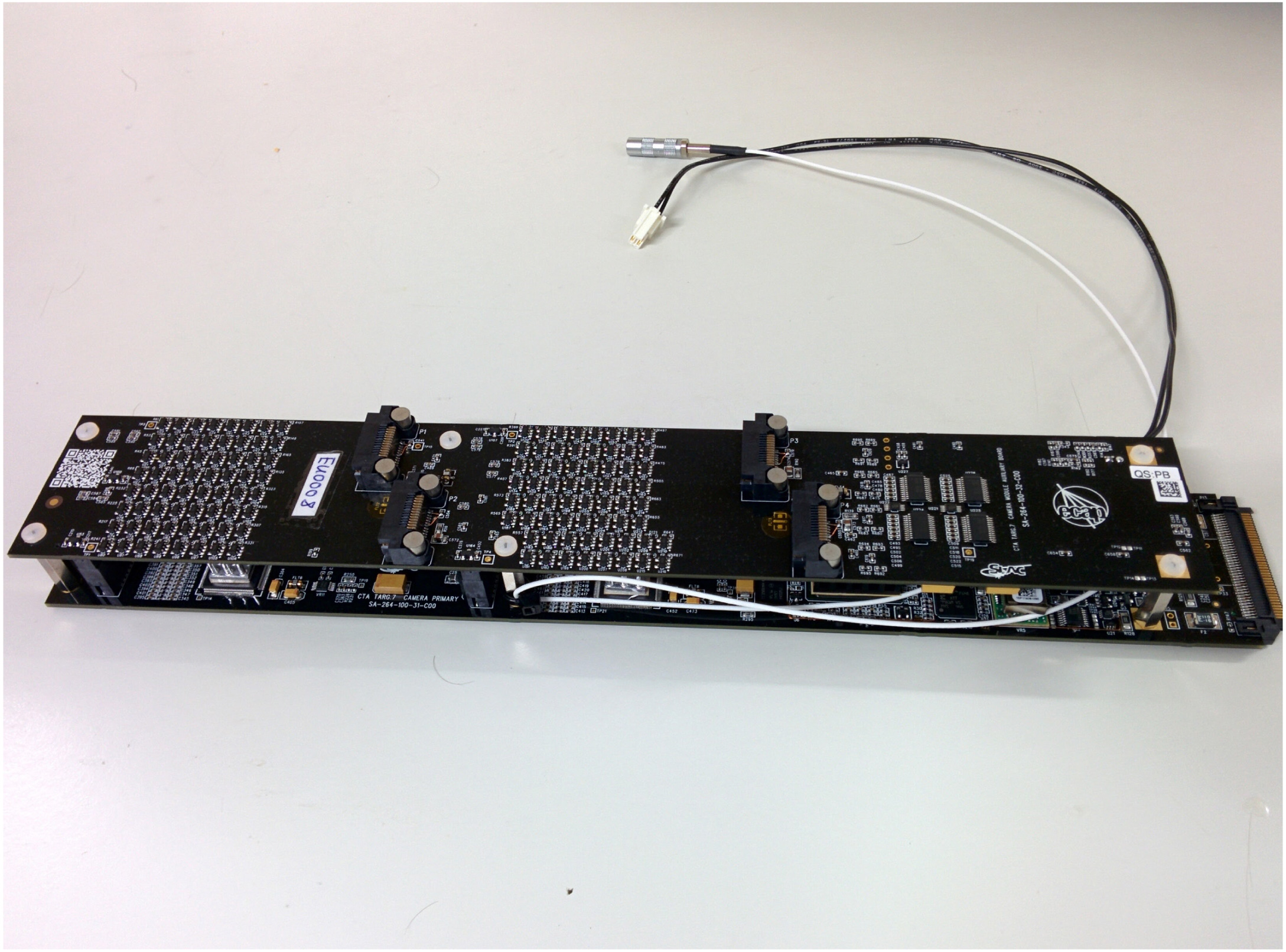}\includegraphics[height=4.5cm]{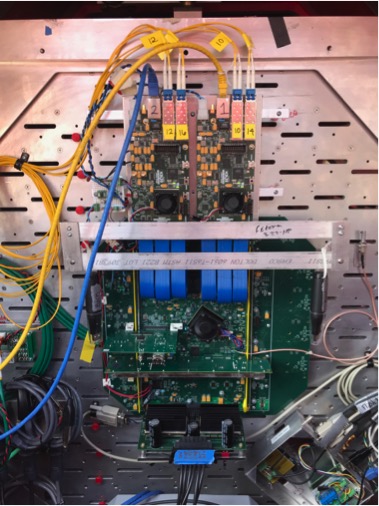}
\caption{Left: Target 7 readout module. Right: Back of SCT camera}
\label{fig:psct2}
\end{figure}

The SCT camera has a modular layout with a total of 11328 pixels assembled into 177 modules.
A module consists thus of 64 =8x8 imaging SiPM pixels.The dimensions of one pixel are 6.5x6.5mm$^2,$ which corresponds to an angular size
of 0.067$^{\mathrm{o}}.$ 
 The focal surface is not a plane so that photon detectors of each
module are adjusted to a different height to approximate the ideal curvature of the SC optics.
The complete readout electronics is integrated in each module. 
The front end electronics associated to a module
is comprised of an auxiliary board and primary board stacked on top of one another (Left panel of Fig. \ref{fig:psct2}).
The auxiliary board performs the pulse shaping and temperature control.
After amplification and shaping, the data are sent into 4 TARGET7 application-specific integrated circuits (ASIC) located on the primary board.
Each TARGET chip serving 16 channels is equipped with a 16,384 cell SCA providing a memory depth of 6.3 $\mu$s at the sampling rate of 1 GHz. 
When a readout command is issued, an effective 10 bit ADC converter digitizes the sample voltages stored in the SCA.
The TARGET chip has also the capacity of performing the first level trigger  by forming the analog sum of four adjacent image pixels and discriminating the summed
signal. 
The camera is divided into 9 sectors, each holding up to 25 modules and a single backplane board.
The backplane receives 16 level 1 (L1) triggers from each module through the primary board
connector.  A FPGA on the backplane provides timing synchronization
and trigger logic. 
Two DAQ boards, mounted on the top of the backplane, are connected by 4 fiber cables to the network switch.
The event data are sent through the DAQ boards to a data server. This data server acts as a local repository 
and is also responsible for slow control and run control.

The pSCT prototype camera has been partially equipped with 24 modules and 2 types of SiPM photodetectors (Right panel of Fig \ref{fig:psct2}). 
As mentionned earlier, the pSCT camera had its first light in 2019.
More details on the pSCT camera performances are given in \cite{bib:Taylor}. The path to the construction of the full size camera is described in \cite{bib:Meures}.
\section{Conclusion and propects}
Two telescope structures and three camera concepts have been developped for the medium sized telescopes. A DC structure prototype has been operated in Adlershof. FlashCam and NectarCAM
prototypes have been mounted on that structure and had their first light in 2017 and 2019, respectively. An SC telescope prototype, the pSCT is being commisionned and had its first light in 2019.
 
\acknowledgments We gratefully acknowledge financial support from the agencies and organizations listed on page:
https://www.cta-observatory.org/consortium\_acknowledgments.

\end{document}